\documentstyle[multicol,aps,pra,epsf]{revtex}

\begin{document}

\title{Random quantum magnets with long-range correlated disorder:\\
Enhancement of critical and Griffiths-McCoy singularities}
 
\author{Heiko Rieger$^{1,2}$ and Ferenc Igl\'oi$^{3,4}$}
\address{
   $^1$ FB 10.1 Theoretische Phyisk, Universit\"at des Saarlandes,
   66041 Saarbr\"ucken, Germany\\
   $^2$ NIC c/o Forschungszentrum J\"ulich, 52425 J\"ulich, Germany\\
   $^3$ Research Institute for Solid State Physics and Optics,   
        H-1525 Budapest, P.O.Box 49, Hungary\\ 
   $^4$ Institute for Theoretical Physics,
        Szeged University, H-6720 Szeged, Hungary}
\date{\today}

\maketitle

\begin{abstract}  
 We study the effect of spatial correlations in the quenched disorder
 on random quantum magnets at and near a quantum critical point. In the
 random transverse field Ising systems disorder correlations that decay
 algebraically with an exponent $\rho$ change the universality class
 of the transition for small enough $\rho$ and the off-critical
 Griffiths-McCoy singularities are enhanced. 
 We present exact results for 1d utilizing a mapping to fractional
 Brownian motion and generalize the predictions for the critical
 exponents and the generalized dynamical exponent in the
 Griffiths phase to $d\ge2$.
\end{abstract}

\pacs{PACS numbers:  75.50.Lk, 71.55.Jv, 75.10.jm, 75.10.Nr}


\newcommand{\bc}{\begin{center}}
\newcommand{\ec}{\end{center}}
\newcommand{\be}{\begin{equation}}
\newcommand{\ee}{\end{equation}}
\newcommand{\beqn}{\begin{eqnarray}}
\newcommand{\eeqn}{\end{eqnarray}}

\begin{multicols}{2}
\narrowtext
\parskip=0cm

The presence of quenched disorder is known to have a pronounced effect
on quantum (or zero-temperature) phase transitions: new universality
classes emerge, some of them with unconventional scaling properties,
quantum activated dynamics and strong, so called Griffiths-McCoy (GM)
singularities \cite{griffiths,mccoy} even away from the critical
point. A well studied example is the random transverse Ising model,
for which many results in one dimension exist
\cite{fisher,mckenzie,big1d}. Many of the unusual properties in 1d
were recently shown to persist also in higher dimensions
\cite{senthil,2dmc,motrunich}. In this paper we investigate the the
effect of long-range correlated disorder on the critical and
off-critical singularities at quantum phase transitions. This issue
has, to our knowledge, never been studied before, although it is known
to have significant impact on various other physical phenomena like
classical (thermal) phase transitions \cite{weinrib}, surface
properties \cite{kpz,scheidl}, anomalous diffusion in disordered
environments\cite{bouchaud} and many other areas.

In addition to the experimental relevance for the quantum Ising spin
glasses \cite{rosen} an interesting relation of the random transverse
Ising models to the non-Fermi liquid behavior of $f$-electron
compounds has recently been revealed \cite{neto}. For this system the
low temperature properties of the interacting Kondo impurities have
been mapped onto an effective quantum spin-$1/2$ system with strong
Ising anisotropy and random bond and transverse fields, which turn out
to possess long range spatial correlations that decay like a power law
with distance \cite{neto,neto2}. This observation, in addition to the
above mentioned general interest, motivates the study of the present
paper, which investigates the effect of long range spatial
correlations in the quenched disorder on the quantum phase transition
as well as on the GM singularities.

To be concrete we consider the Ising ferromagnet with
transverse fields
\be
H=-\sum_{\langle ij\rangle} J_{ij}\sigma_i^x\sigma_j^x-
\sum_i h_i\sigma_i^z\;,
\label{hamil}
\ee
where $\sigma_i^{x,z}$ are spin-$1/2$ operators and the interactions
$J_{ij}$ ($\ge0$) and/or the transverse fields $h_i$ ($\ge0$) are
quenched random variables. The spins are located on the sites of a
$d$-dimensional lattice and $\langle ij\rangle$ denote nearest
neighbor pairs on this lattice (modeling short-range interactions).

At zero temperature ($T=0$) this model has, in any dimension $d$ a
quantum phase transition from a paramagnetic to a ferromagnetic phase
at some critical value of the average ratio between bonds and fields
$[ln (h_i/J_{ij})]_{\rm av}$. Here and in the following $[\ldots]_{\rm
  av}$ denotes the disorder average. The distance from this critical
point is $\delta$, such that for $\delta>0$ the system (\ref{hamil})
is paramagnetic, for $\delta<0$ it is ferromagnetic. We can introduce
local deviations from the critical point by introducing variables
$\delta({\bf r})$, which in 1d are simply given by
$\delta(r)=\ln(h_r/J_{r,r+1})$.  Thus one can discriminate regions in
space that tend {\it locally} to be ferromagnetic (paramagnetic) even
for $\delta>0$ ($\delta<0$).

Here we wish to study systematically the effect of (isotropic) spatial
correlations in the disorder that can be modeled with a disorder
correlator $G({\bf r})$:
\be
[\delta({\bf r})\delta({\bf r'})]_{\rm av}=G({\bf r}-{\bf r'})\;.
\label{corr}
\ee
Uncorrelated disorder is described by choosing $G({\bf r})$ to be a
delta-function. The Harris criterion for correlated disorder
\cite{weinrib,weinrib_perc} shows that any disorder correlator that
falls off faster than $r^{-2/\nu}$ (i.e.\ $G({\bf r})\sim{\cal
O}(r^{-\rho})$ with $\rho>2/\nu$, where $\nu$ is the correlation
length exponent for uncorrelated disorder) does not change the
universality class of the quantum critical point of model
(\ref{hamil}) with uncorrelated disorder.
On the other hand for 
\be
G({\bf r})\sim r^{-\rho}\quad{\rm with}\quad0<\rho\le 2/\nu\quad(\le d)\,;
\ee
where the last inequality holds generally for disordered system
with uncorrelated disorder \cite{chayes} (c.f. \cite{pazmandi}), the
disorder correlations are relevant (and thus truly long-ranged), the
critical exponents become different from the uncorrelated case and the
quantum critical point constitutes a new universality class, which we
are going to explore in this paper.

First we consider the 1d case for which we can derive most of our
results in a rigorous way utilizing a mapping of the problem to
fractional Brownian motion \cite{mandelbrot,feder}. First we note that
$\nu=2$ for uncorrelated disorder \cite{fisher,big1d}, which means
that the long-range correlations are relevant for $\rho<1$. The
critical exponents and scaling relations can be determined by studying
the finite size scaling behavior of model (\ref{hamil}). As it is
shown in \cite{big1d} the gap $\Delta E$ (lowest excitation energy) of
a chain of length $L$ is given by $\Delta E\approx m_s \overline{m}_s
h_L \prod_{i=1}^{L-1} h_i/J_{i,i+1}$, where $m_s$ and $\overline{m}_s$ are
the left and right surface magnetizations ($\sim{\cal O}(1)$),
respectively. Thus,
\be
\ln\Delta E\propto{\textstyle\sum_{i=1}^{L-1}}\delta(i)\;.
\label{gap}
\ee
Since $\delta(i)$ are random variables with zero mean at criticality
that are correlated according to (\ref{corr}) we conclude that
$\ln\Delta E$ scales like the transverse fluctuations of a correlated
random walk. Since $[\{\sum_{i=1}^{L-1}\delta(i)\}^2]_{\rm av} \sim
L\int_1^L dr\,G(r)\approx L^{2-\rho}$ we have for $\rho\le1$
\be
\ln\Delta E\propto L^{\psi(\rho)}\quad{\rm with}\quad
\psi(\rho)=1-\rho/2\;.\label{energyscale}
\ee
Thus for long-range-correlated disorder the quantum activated dynamics
scenario at the critical point is even enhanced and we get a new
critical exponent $\psi$. Note that for $\rho\ge1$ one gets
$\psi=1/2$, the result for uncorrelated disorder
\cite{fisher,big1d}. We checked this result by computing numerically
the probability distribution $P_L(\Delta E)$ (see \cite{big1d} for
details), which we indeed found to scale like $P_L(\ln\Delta E)\sim
L^{-\psi(\rho)}\tilde{p}(\ln\Delta E/L^{\psi(\rho)})$ with $\psi(\rho)$
as in (\ref{energyscale}).

The surface magnetization $m_s=\langle\sigma_1^x\rangle$ of a finite
chain of length $L$ (with the spin at site $L$ being fixed) is given
by $m_s=\{1+\sum_{k=1}^{L-1}\prod_{i=1}^k
(h_i/J_{i,i+1})^2)\}^{-1/2}$, \cite{big1d}. From this expression
and (\ref{energyscale}) one sees that $[\ln m_s(L)]_{\rm av}\sim
-L^{\psi(\rho)}$, i.e.\ that the {\it typical} magnetization decays
with a stretched exponential. Moreover, away from the critical point
($\delta>0)$ one has $[\ln m_s(L)]_{\rm av}\sim -L\delta$ implying
$[m_s(L,\delta)]_{\rm typ}\propto\exp(-L/\xi_{\rm typ})$, where we
defined {\it typical} correlation length that is seen to scale like
$\xi_{\rm typ}\sim\delta^{-\nu_{\rm typ}}$ with $\nu_{\rm typ}=1$
independent of the correlation exponent $\rho$.

On the other hand the {\it average} surface magnetization can be shown
\cite{big1d,diffusion} to scale like the survival probability $P_{\rm surv}(L)$
of a random walk of $L$ steps. This can be related to the first return
time distribution $P_{\rm f.r.t.}(L)$ of a fractional Brownian motion
(with Hurst exponent $H=1-\rho/2\in[1/2,1]$,
c.f. (\ref{gap},\ref{energyscale})), which has been shown to scale
like $L^{H-2}$ \cite{dingyang}. Since $P_{\rm surv}(L)=\int_L^\infty
P_{\rm f.r.t.}(L)\sim L^{H-1}$ we get for $\rho\le1$
\be
[m_s(L)]_{\rm av}\sim L^{-x_s(\rho)}\quad{\rm with}\quad
x_s(\rho)=\rho/2
\label{surfexp}
\ee
For $\rho\ge1$ one has the known result $x_s=\beta_s/\nu=1/2$. 

From the analogy to fractional Brownian motion one can also derive the
exponent $\nu$ describing the divergence of the {\it average}
correlation length when approaching the critical point,
$\xi\sim|\delta|^{-\nu}$. A non-vanishing distance from the critical
point implies that the disorder configurations are such that they give
rise to a non-vanishing average for the step width $[\delta(i)]_{\rm
  av}=\delta$, i.e.\ the fractional Brownian motion is {\it biased}.
For $\delta>0$ (in the paramagnetic phase) the return time
distribution has an exponential cut-off beyond a characteristic length
scale $\xi$ that scales like $\delta^{1/H-1}$ \cite{dingyang}, which
yields for $\rho\le1$
\be
\xi_{\rm av}\sim\delta^{-\nu(\rho)}
\quad{\rm with}\quad\nu(\rho)=2/\rho\;.
\label{corrlength}
\ee
From (\ref{surfexp}) and (\ref{corrlength}) one gets
$\beta_s(\rho)=1$, independent of $\rho$.  For $\rho\ge1$ it is
$\nu=2$. The finite size scaling behavior of the average surface
magnetization is then described by $[m_s]_{\rm
  av}=L^{-x_s}\tilde{m}(L^{1/\nu}\delta)$.  In Fig. 1 we show a
corresponding scaling plot for $\rho=0.75$.

\begin{figure}[t]
\epsfxsize=\columnwidth\epsfbox{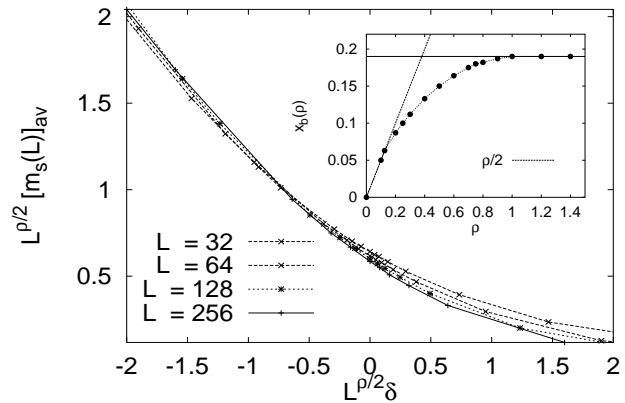}
\caption{Finite size scaling plot of the surface magnetization 
  according to the form $[m_s]_{\rm av} = L^{-x_s} \tilde{m}
  (L^{1/\nu}\delta)$ for $\rho=0.75$ using (\ref{corrlength})
  and (\ref{surfexp}). The data are averaged over 50000 samples using
  a symmetric binary distribution for the couplings
  ($J_{i,i+1}\in\{\lambda,\lambda^{-1}\}$, here
  $\lambda=5$, $h_i=1$).\\ {\bf Inset:} The bulk magnetization exponent
  $x_b$ as a function of the value of disorder correlation exponent
  $\rho$ estimated by evaluating numerically the average persistence
  exponent for the same type of correlated disorder in the Sinai-model
  (see text). The horizontal line is the value for uncorrelated disorder
  $x_b=(3-\protect{\sqrt{5}})/4=0.19098\ldots$, the dashed line is
  $\rho/2$ and represents the asymptotic dependence of $x_b(\rho)$ for
  $\rho\to0$.
}
\end{figure}

The bulk magnetization $m_b=\langle\sigma_{L/2}^x\rangle$ of such a
chain is much harder to calculate, see \cite{big1d}. The size
dependence of the average bulk magnetization at criticality determines
the last and remaining critical exponent $x_b$ via
\be
[m_b(L)]_{\rm av}\sim L^{-x_b(\rho)}
\label{bulkmag}
\ee
In the case of uncorrelated disorder it was possible to predict the
exact bulk magnetization exponent $x_b=\beta_b/\nu=(3-\sqrt{5})/4$
(that will also hold for any $\rho\ge1$) using a particular real space
renormalization group \cite{fisher}, which, however, appears to be
inappropriate for long-range correlated disorder. In the limit
$\rho\to0$ the difference between bulk and surface magnetization will
vanish due to the extreme correlation of the disorder. Hence we expect
$x_b(\rho)\approx x_s(\rho)=\rho/2$ for $\rho\to0$. To obtain the full
$\rho$-dependence we computed numerically the average magnetization
for finite systems using the fermion representation \cite{big1d} and
analyzed its finite size scaling behavior. We found only a slight
decrease of $x_b(\rho)$ with $\rho$, which presumably does not reflect
the true asymptotic behavior that is much harder to reach numerically
for correlated disorder than for uncorrelated disorder. Therefore we
used the relation between the scaling properties of the bulk
magnetization and the average persistence of a Sinai walker
\cite{diffusion,sinai} to compute the exponent $x_b(\rho)$, as shown
in the inset of Fig.  1. It confirms the asymptotic behavior
$x_b(\rho)\sim\rho/2$ and the expected inequality $x_b(\rho)\le
x_s(\rho)$ \cite{sinai}. In passing we note that the order parameter
profiles $m(r)=[\langle\sigma_{r}^x\rangle]_{\rm av}$ do not have the
simple scaling properties reported in \cite{profiles} for the
uncorrelated case.

Away from the critical point the physical properties are controlled by
strongly coupled clusters (i.e.\ segments that have locally a tendency
to order ferromagnetically \cite{big1d}) giving rise to the so called
GM singularities. In the paramagnetic phase ($\delta>0$)
the probability $P_L(l)$ for a strongly coupled cluster of length $l$
is proportional to $Le^{-l/\xi}$ implying a typical length of such a
cluster $l_{\rm typ}\sim\xi\ln L$, with $\xi$ given by
(\ref{corrlength}). With eq(\ref{gap}) for the gap we therefore get a
typical energy scale of $\ln\Delta E\propto \delta l_{\rm typ}
\propto-\delta^{1-\nu(\rho)}\ln L$. Therefore we obtain for $\rho\le1$
and $\delta\ll1$
\be
\Delta E\sim L^{-z'(\delta,\rho)}\quad{\rm with}\quad
z'(\delta,\rho)\propto\delta^{1-2/\rho}
\label{zgriff}
\ee
(and $z'(\rho,\delta)=2\delta^{-1}$ for $\rho\ge1$, $\delta\ll1$). The
generalized dynamical exponent $z'(\rho,\delta)$ parameterizes {\it
  all} singularities occurring in the GM phase
\cite{fisher,big1d,griffsing}: e.g.\ the spin autocorrelation function
at $T=0$ decays algebraically with $z'$, $G_{\rm
  loc}(\tau)=[\langle\sigma_i^x(\tau)\sigma_i(0)\rangle]_{\rm
  av}\sim\tau^{-1/z'}$; the local susceptibility diverges for $T\to0$
when $z'>1$, $\chi_{\rm loc}\sim T^{1/z'-1}$; the specific heat has an
algebraic singularity at $T=0$, $C\sim T^{1/z'}$; the magnetization in
the presence of an external longitudinal field (in the $x$-direction)
scales as $M\sim H^{1/z'}$ etc. We computed numerically \cite{big1d}
the probability distribution $P_L(\Delta E)$ which we confirmed to
have a power law tail $P_L(\Delta E)\propto\Delta E^{-1/z'+1}$ with
$z'$ as in (\ref{zgriff}).

Concluding 1d case we stress that (for fixed distance $\delta$ from
the critical point) $z'$ increases monotonically with decreasing
$\delta$, i.e.\ stronger disorder correlations generate stronger
GM singularities. Due to the nature of these
singularities this tendency is a direct consequence of an increasing
probability for large clusters for increasing disorder correlations.

Before we proceed to the higher dimensional case ($d\ge2$) we describe
briefly the infinite randomness disorder fixed point (IRFP) scenario,
originally developed for uncorrelated disorder\cite{fisher2}, but one
can generalize it for the present correlated case. This
phenomenological theory involves three exponents: the lowest energy
scale $\Delta E$ and the linear size $L$ of a strongly coupled cluster
are related via $\ln\Delta E\sim L^\psi$, its magnetic moment scales
as $\mu\sim L^{\phi\psi}$ and its typical size when approaching the
critical point, the correlation length, will diverge like
$\xi\sim|\delta|^{-\nu}$.  All bulk exponents can be expressed via
$\psi,\phi$ and $\nu$, c.f. $x_b=d-\phi\psi$, $\nu_{\rm
  typ}=\nu(1-\psi)$ and in the Griffiths phase
$z'\propto\delta^{-\nu\psi}$. For the 1d case, as treated above, it is
$\psi=1/2$, $\phi=(\sqrt{5}+1)/2$ and $\nu=2$ for uncorrelated
disorder and $\rho>1$, whereas for $\rho<1$ we obtained
$\psi=1-\rho/2$, $\phi=(1-x_b)/\psi$ (see Fig. 1) and $\nu=2/\rho$.
The exponent relations are satisfied for the correlated and
uncorrelated cases.

In higher dimensions the IRFP scenario still holds
\cite{fisher2,motrunich}, irrespective of the presence or absence of
disorder correlations. However, the exponents ($\psi,\phi,\nu$) will
change. We can make precise statements on the change of these
exponents for the case of random bond or site dilution, for which the
quantum phase transition occurs at the percolation threshold $p=p_c$
(with $p$ being the bond- or site concentration).  In this case the
physics is completely determined by the {\it geometric} properties of
the percolating clusters \cite{senthil}, which means that
($\psi,\phi,\nu$) can be expressed by the classical percolation
exponents, which are the fractal dimension $D_{\rm perc}$ of the
percolating cluster, the exponent $\beta_{\rm perc}$ determining the
probability for a site being in the percolating cluster, and the
correlation length exponent $\nu_{\rm perc}$, respectively. It is known
\cite{weinrib,weinrib_perc} that the disorder correlations are
relevant for $\rho<2/\nu_{\rm perc}$, in which case one gets
$\nu(\rho)=\nu_{\rm perc}(\rho)=2/\rho$, $\psi(\rho)=D_{\rm
  perc}(\rho)$, which will increase with increasing correlations,
since then clusters become more compact, and $\phi(\rho)=(d-\beta_{\rm
  perc}(\rho)/\nu_{\rm perc}(\rho))/D_{\rm perc}(\rho)$
where $\beta_{\rm
  perc}(\rho)/\nu_{\rm perc}(\rho)$ is possibly independent of $\rho$
\cite{stanley}. In the GM phase this implies for
the dynamical exponent
\be
z'\sim|p-p_c(\rho)|^{-2D_{\rm perc}/\rho}\;,
\ee
which increases with decreasing $\rho$, again confirming the
general tendency that disorder correlations enhance the
GM singularities.

For the generic non-diluted case the exponents need not to be
identical with the diluted case, however, one still has $\nu=2/\rho$
for $\rho<2/\nu_{\rm uncorr}$, according to a general argument given
in \cite{weinrib}. Moreover, $\psi$ increases with increasing disorder
correlations, since its value is connected to the geometric
compactness of strongly coupled clusters. Thus, the dynamical exponent
$z' \sim \delta^{-\nu\psi}$ grows, again enhancing the GM singularities.

Regarding the recent experiments on $f$-electron systems we would like
to point out that there is evidence \cite{neto2} that the spatial
correlations in the metallic compound U$_{1-x}$Th$_x$Pd$_3$ decay like
$r^{-3}$. If we assume that for the transverse Ising model with
uncorrelated disorder we have $\nu=2/d$ (as it is the case for d=1
\cite{fisher,big1d} and d=2 \cite{motrunich}, and also for other
random quantum critical points \cite{qsg,bg}) such a decay with
$\rho=3$ is the marginal case and instead of modifications of the
above quoted exponents strong logarithmic corrections will appear.  In
a scaling theory for the marginal situation one has to replace $L$ and
$\xi$ by $L\ln L$ and $\xi\ln\xi$, respectively yielding for $d$=$3$
and $\rho$=$3$ for the gap, correlation length and dynamical exponent
in the GM phase
\beqn
\ln \Delta E & \sim & L^\psi\ln^{\psi}L\;, \nonumber\\
\xi          & \sim & \delta^{-\nu} |\ln\delta|^{-1}\;,\\
z'           & \sim & \delta^{-\nu\psi}|\ln\delta|^{\psi}\;, \nonumber
\eeqn
respectively, where $\psi$ and $\nu$ are the critical exponents of the
3d system with {\it uncorrelated} disorder. Obviously these logarithmic
corrections will make it very hard to extract the critical exponents
$\psi$ and $\nu$ for instance from experimental data, and,
furthermore, will {\it apparently} vary when approaching the critical
point.

Finally we would like to make a few remarks on related quantum
magnetic systems: a) Quantum spin glasses \cite{qsg} are expected to
behave very similar as the random ferromagnets with respect to the
introduction of disorder correlations. The frustration caused by the
random signs of the spin interactions is irrelevant at the IRFP
\cite{fisher2} and therefore the universality class and the critical
exponents are not changed.  b) Random XY, e.g.\ in 1d
\be
H=\sum_i  J_{i,i+1}^x\sigma_i^x\sigma_{i+1}^x
         +J_{i,i+1}^y\sigma_i^y\sigma_{i+1}^y
\ee
and XXZ or Heisenberg systems have different features in 1d and in
d$\ge2$. In the latter higher dimensional case it seems that the
quantum critical point is {\it not} an IRFP \cite{fisher2,takayama},
however, disorder correlations will certainly affect the critical
properties. In 1d we encounter the same scenario as for the transverse
Ising systems, in particular for XY and XX chains \cite{xy}, since
these are equivalent to two decoupled transverse Ising chains. Most
remarkably the transverse and longitudinal correlations still decay
with the same exponent (in contrast to the pure case), however, more
slowly with correlated disorder. Moreover the term {\it random singlet
  phase} is now inapproriate when $\rho<1$, since then larger units
than only pairs of spins will be strongly coupled.

To summarize we have studied the effect of long-range correlations in
the disorder on the quantum critical behavior of random magnets. We
have shown the relevant correlations generally enhance the critical
and off-critical singularities, essentially because large strongly
coupled clusters appear more frequently. For the random transverse
Ising system in 1d we reported exact values for the critical exponents
for arbitrary disorder correlation exponent $\rho$, also for the
diluted case in higher dimensions and argued how these results can be
generalized to generic bond/field randomness in $d\ge2$. With respect
to the recent experiments on f-electron systems we have pointed out
the existence of strong logarithmic corrections that complicates the
measurement of the critical and the dynamical exponenent. Finally we
generalized our results to $XX$ and $XY$ quantum spin systems.

F.\ I.'s work has been supported by the Hungarian National Research
Fund under grant No OTKA TO23642, TO25139, MO28418 and by the
Ministery of Education under grant No. FKFP 0596/1999.  H.\ R. was
supported by the Deutsche Forschungsgemeinschaft (DFG).

\end{multicols}

\begin{references}

\bibitem{griffiths} 
\vskip-1.3cm
R.~B.~Griffiths, Phys. Rev. Lett. {\bf 23}. 17
(1969).

\bibitem{mccoy}
B.~M.~McCoy, Phys. Rev. Lett. {\bf 23}, 383 (1969); Phys. Rev. {\bf
188}, 1014 (1969).

\bibitem{fisher}
D.~S.~Fisher, Phys. Rev. Lett. {\bf 69}, 534 (1992); Phys. Rev. B {\bf
51}, 6411 (1995).

\bibitem{mckenzie}
R. H. McKenzie, 
Phys. Rev. Lett. {\bf 77}, 4804 (1996).

\bibitem{big1d}
F. Igl\'oi and H.\ Rieger,
Phys. Rev. B {\bf 57}, 11404 (1998).

\bibitem{senthil}
T.~Senthil and S.~Sachdev, Phys. Rev. Lett {\bf 77}, 5292 (1996).

\bibitem{2dmc}
C. Pich, A. P. Young, H. Rieger, and N. Kawashima, Phys. Rev. Lett.
{\bf 81}, 5916 (1998); H. Rieger and N. Kawashima, Europ. Phys.
J. B {\bf 9}, 233 (1999); T. Ikegami, S. Miyashita and H. Rieger,
J. Phys. Soc. Jap. {\bf 67}, 2761 (1998).
  
\bibitem{motrunich}
O. Motrunich, S.-C. Mau, D. A. Huse, D. S. Fisher, 
cond-mat/9906322.

\bibitem{weinrib}
A.\ Weinrib and B. I. Halperin,
Phys. Rev. B {\bf 27}, 413 (1983).

\bibitem{kpz}
M. Kardar, G. Parisi and Y.-C. Zhang, 
Phys. Rev. Lett. {\bf 56}, 889 (1986);
T. Halpin-Healy and Y.-C. Zhang, 
Phys. Rep. {\bf 254}, 215 (1995) and references therein.

\bibitem{scheidl}
S. Scheidl, Phys. Rev. Lett. {\bf 75}, 4760 (1995). 

\bibitem{bouchaud}
J.\ P.\ Bouchaud and A.\ Georges, 
Phys.\ Rep.\ {\bf 195}, 127 (1990) and references therein.

\bibitem{rosen}
W. Wu, et al.,
Phys. Rev. Lett. {\bf 67}, 2076 (1991);
W. Wu, et al.,
Phys. Rev. Lett. {\bf 71}, 1919 (1993).

\bibitem{neto}
A. H. Castro Neto, G. Castilla, and B. Jones, 
Phys. Rev. Lett. {\bf 81}, 3531 (1998);
M. C. Andrade et al., 
Phys. Rev. Lett. {\bf 81}, 5620 (1998).

\bibitem{neto2}
A. H. Castro Neto, Private communication.

\bibitem{weinrib_perc} 
A.\ Weinrib, Phys. Rev. B {\bf 29}, 387 (1984).


\bibitem{chayes} 
J. T. Chayes, L. Chayes, D. S. Fisher and T. Spencer,
Phys. Rev. Lett. {\bf 57}, 2999 (1986).

\bibitem{pazmandi}
F. P\'azm\'andi, G. T. Zim\'anyi and R. Scalettar,
Phys. Rev. Lett. {\bf 79}, 5190 (1997).

\bibitem{mandelbrot}
B. B. Mandelbrot and J. W. van Ness,
SIAM Rev. {\bf 10}, 422 (1968).

\bibitem{feder}
J. Feder: {\it Fractals} (Plenum Press, NewYork-London, 1988).

\bibitem{diffusion}
F. Igl\'oi and H.\ Rieger, 
Phys. Rev. E {\bf 58}, 4238 (1998).

\bibitem{dingyang}
M. Ding and W. Yang,
Phys. Rev. E {\bf 52}, 207 (1995).

\bibitem{sinai}
H.\ Rieger and F. Igl\'oi,
Europhys. Lett. {\bf 45}, 673 (1999).

\bibitem{profiles}
F. Igl\'oi and H.\ Rieger,
Phys. Rev. Lett. {\bf 78}, 2473 (1997).

\bibitem{griffsing}
M. J. Thill and D. A. Huse, 
Physica A {\bf 15}, 321 (1995);
H.\ Rieger and A.\ P.\ Young,
Phys.\ Rev.\ B {\bf 54}, 3328 (1996);
F.\ Igl\'oi, R. Juh\'asz, and H.\ Rieger,
Phys. Rev. B {\bf 59}, 11308 (1999).

\bibitem{stanley}
S. Prakash, S. Havlin, M. Schwartz, and H. E. Stanley,
Phys. Rev. A {\bf 46}, R1724 (1992).

\bibitem{fisher2}
D. S. Fisher,
Physica A {\bf 263}, 222 (1999).

\bibitem{qsg}
H. Rieger and A. P. Young, 
Phys. Rev. Lett. {\bf 72}, 4141 (1994);
M. Guo, R. N. Bhatt and D. A. Huse, 
Phys. Rev. Lett. {\bf 72}, 4137 (1994).

\bibitem{bg}
E. S. S\o rensen, M. Wallin, S. M. Girvin and A. P Young,
Phys. Rev. Lett. {\bf 69}, 828 (1992);
J.\ Kisker and H.\ Rieger,
Phys.~Rev.\ B {\bf 55}, 11981R (1997).

\bibitem{takayama}
K. Kato, et al. cond-mat/9905379.

\bibitem{xy}
H. Rieger, R. Juh\'asz and F. Igl\'oi,
cond-mat/9906378; and to be published.


\end{references}
\end{document}